 \documentclass[preprint2,longabstract]{aastex}

\usepackage{multirow}

\shorttitle{TBA}
\shortauthors{Maltseva et al.}

\begin{document}

\title{High-resolution IR absorption spectroscopy of polycyclic aromatic hydrocarbons: the realm of anharmonicity}

\author{Elena Maltseva}
\affil{University of Amsterdam, Science Park 904, 1098 XH Amsterdam, The Netherlands}
\and
\author{Annemieke Petrignani\altaffilmark{1}, Alessandra Candian, Cameron J. Mackie}
\affil{Leiden Observatory, Niels Bohrweg 2, 2333 CA Leiden, The Netherlands}
\email{petrignani@strw.leidenuniv.nl}
\altaffiltext{1}{Radboud University, Toernooiveld 7, 6525 ED Nijmegen, The Netherlands}
\and
\author{Xinchuan Huang}
\affil{SETI Institute, 189 Bernardo Avenue, Suite 100, Mountain View, California 94043, USA}
\and
\author{Timothy J. Lee}
\affil{NASA Ames Research Center, Moffett Field, CA 94035-1000, USA}
\and
\author{Alexander G. G. M. Tielens}
\affil{Leiden Observatory, Niels Bohrweg 2, 2333 CA Leiden, The Netherlands}
\and
\author{Jos Oomens}
\affil{Radboud University, Toernooiveld 7, 6525 ED Nijmegen, The Netherlands}
\and
\author{Wybren Jan Buma}
\affil{University of Amsterdam, Science Park 904, 1098 XH Amsterdam, The Netherlands}
\email{w.j.buma@uva.nl}

\begin{abstract}
We report on an experimental and theoretical investigation of the importance of anharmonicity in the 3-$\mu$m CH stretching region of Polycyclic Aromatic Hydrocarbon (PAH) molecules. We present mass-resolved, high-resolution spectra of the gas-phase cold ($\sim$4K) linear PAH molecules naphthalene, anthracene, and tetracene. The measured IR spectra show a surprisingly high number of strong vibrational bands. For naphthalene, the observed bands are well separated and limited by the rotational contour, revealing the band symmetries. Comparisons are made to the harmonic and anharmonic approaches of the widely used Gaussian software. We also present calculated spectra of these acenes using the computational program SPECTRO, providing anharmonic predictions with a Fermi-resonance treatment that utilises intensity redistribution. We demonstrate that the anharmonicity of the investigated acenes is strong, dominated by Fermi resonances between the fundamental and double combination bands, with triple combination bands as possible candidates to resolve remaining discrepancies. The anharmonic spectra as calculated with SPECTRO lead to predictions of the main modes that fall within 0.5\% of the experimental frequencies. The implications for the Aromatic Infrared Bands, specifically the 3-$\mu$m band are discussed.
\end{abstract}


\section{Introduction}
Polycyclic Aromatic Hydrocarbon (PAH) species are considered to be responsible for the family of infrared (IR) emission features, the so-called Aromatic Infrared Bands (AIBs), that dominate the spectrum of objects ranging from protoplanetary disks to entire galaxies (see \citet[]{JT11}, and reference therein). PAH molecules absorb visible to ultraviolet (UV) photons and then relax through IR emission \citep[]{Sell84, ATB89, Puge89}. These spectral signatures are due to vibrational modes that are typical of a family of molecules showing the same chemical bond types.

Over the years, a wealth of theoretical (see e.g. the Cagliari and NASA AMES databases \citep{Mall07,Boer11,Boer14b}) and experimental (see e.g. \citet{Oome11}, and reference therein) studies of the vibrational spectrum of different types of PAH molecules (neutral, charged, heteroatom substituted, etc.) have been reported with the aim to aid the identification of subclasses of PAHs when compared to the global characteristics of the AIBs in well-known astronomical sources \citep{Cand12,Boer12,Boer13,Boer14a}.

Both theoretical and experimental studies have known caveats. Apart from rare exceptions \citep{Cane07,Pira09,Mari15}, Density Functional Theory (DFT) calculations have been performed exclusively with the double harmonic approximation, thus neglecting the effects of anharmonicity. The calculations by \citet{Pira09} have been the most extensive in incorporating the effects of anharmonicity, but have been restricted to naphthalene. Moreover, \citet{Pira09} focused on addressing the detailed profiles of the lower frequency, IR active modes and, in particular, assessing the relative contribution of anharmonicity effects (e.g., hot bands) and rotational structure to measured emission profiles. More specifically \citet{Pira09} considered Fermi (and Darling-Dennison) resonances. What was not included was derivatives beyond the first order of the dipole moment. In contrast, here we present the first results of an extensive study on the effects of anharmonicity on the infrared spectrum of PAHs, combining high-resolution absorption spectra with theoretical studies which allow for anharmonicity as well as Fermi resonances. As the effects of anharmonicity and Fermi resonances are expected to be largest for the CH stretching modes, we focus here on the 3000 cm$^{-1}$ region. In addition, unlike almost all previous gas-phase studies $-$ with the notable exception of \citet{Hune04} $-$, the spectra are measured at very low temperature in order to limit effects of rotational broadening that may obstruct observation of individual bands and to enable an easier and unambiguous assignment of bands. Under such conditions we also avoid (mode-specific) perturbations by the matrix environment that are difficult to predict but inevitably present in matrix-isolation spectroscopy (MIS) studies. As will be shown in the present study, these aspects (resolution and isolated molecules) are key to assess the quality of the calculations and fine-tune the computational approach. 

Commonly, correction factors of about 0.96 \citep{Lang96} are applied to calculated harmonic frequencies to bring them in line with peak positions measured using MIS techniques. As the latter are affected by ill-understood matrix effects, proper validation of the theoretical spectra against laboratory spectra of cold, gas phase PAHs are imperative before they can be used in analysis studies of observed interstellar spectra. To properly account for their astrophysical properties, supersonic molecular beam methods in combination with laser spectroscopy provide an attractive way to study the vibrational spectrum of very cold PAH molecules under isolated conditions at a resolution that is only determined by the laser band width. Recent advances in computational chemistry \citep{BB12} enable efficient anharmonic vibrational analyses of medium sized molecules, providing theoretical tools to properly interpret the experimental results.

In the present study we apply IR-UV double resonance spectroscopy to record mass-selected IR absorption spectra in the 3~$\mu$m region of three small linear PAH molecules: naphthalene, anthracene, and tetracene. We show that these spectra are at odds with harmonic and anharmonic calculations as employed so far, but are in good agreement with our new anharmonic spectra calculated with the program SPECTRO that properly incorporates multiple resonances. 

\section{Methods}

\subsection{Experimental}
IR absorption spectra were recorded using an IR-UV depletion scheme. The molecular beam setup employed for these measurements has been described in detail before \citep{Smol11}. Briefly,  a heated supersonic pulsed source with a typical pulse duration of 190~$\mu$s and argon at a backing pressure of 2.2~bar as a buffer gas was used to create a cold molecular beam. A constant mass-selected ion signal was created via a two-color Resonant Enhanced MultiPhoton Ionization (REMPI) scheme using a Nd:YAG pumped frequency-doubled dye laser to excite the molecules to the first excited state and then an ArF excimer laser to ionize the excited molecules. The resonant excitation in combination with the supersonic cooling conditions ensures that all of bands start from the vibrationless ground state. The IR light in 3-$\mu$m region produced by a Nd:YAG pumped frequency-mixed dye laser with a line width of 0.07 cm$^{-1}$ preceded these two laser beams by 200~ns, leading to dips in the ion signal upon IR absorption. The mass-resolved nature of our experiments and the use of supersonically-cooled molecules ensures that in our experiments we can unambiguously conclude that all of the measured absorption features concern excitations from the vibrational ground state of each PAH, and that they do not derive from isotopically-substituted species.

\subsection{Computational details}
We employed three computational methods; the standard harmonic and anharmonic vibrational approaches, both using Gaussian 09 \citep{G09}, and an anharmonic method utilizing a modified version \citep{Mart95} of the SPECTRO program \citep{Will90, Gaw96} that incorporates intensity sharing due to Fermi resonances \citep{Mack15b}, referred to in this work as G09-h, G09-anh, and SP15 respectively. All calculations apply DFT using a similar integration grid as in \citet{Boes04}, the B9-71 functional \citep{Hamp98} and the TZ2P basis set \citep{Dunn71} that provide the best performance on organic molecules \citep{Boes04, Cane07}. SP15 takes the G09-h intensities and G09-anh force constants and implements its own vibrational second-order perturbation method \citep{Mack15a,Mack15b}. It treats couplings between multiple resonances (states falling within 200 cm$^{-1}$ of each other) simultaneously and redistributes intensity among the modes. Traditionally, two vibrational states of the same symmetry with energies close to each other may interact. The resonance types included in our 2nd-order perturbation theory treatment are beyond the traditional Fermi I and II type, and also include 1-1 and 2-2 type resonances. The coupling terms involve both cubic and quartic anharmonic force constants, and the explicit formula for the 1-1 and 2-2 type resonances are given in \citet{Lehm88,Lehm93}. The implementation of this polyad approach into SPECTRO is described in \citet{Mart95}. No intensity resonances are considered in this study, as we use the Gaussian double-harmonic intensities for intensity distributions.

\section{Results and Discussion}
All IR absorption spectra were recorded between 3.17 and 3.4 $\mu$m (2950 and 3150 cm$^{-1}$). With the present signal-to-noise ratios (S/Ns), no other IR bands were observed outside the displayed range. The respective ion signals were created by fixing the frequency of the first UV photon to the 0-0 band of the S$_1\leftarrow$S$_0$ electronic transition of the respective PAHs. The exact frequencies used are 32028.18~cm$^{-1}$, 27697.0~cm$^{-1}$, and 22402.43~cm$^{-1}$ for naphthalene, anthracene, and tetracene, respectively. These were determined by scanning a small range around the previously reported values from \citet{Hira85,Lamb84,Zhan08} for the S$_1$($^1$B$_{3u})\leftarrow$S$_0$($^1$A$_g$), S$_1$($^1$B$_{1u})\leftarrow$S$_0$($^1$A$_g$), and S$_1$($^1$B$_{2u})\leftarrow$S$_0$($^1$A$_g$) transitions, respectively.

\subsection{Naphthalene (C$_{10}$H$_8$)}

\begin{figure}
\plotone{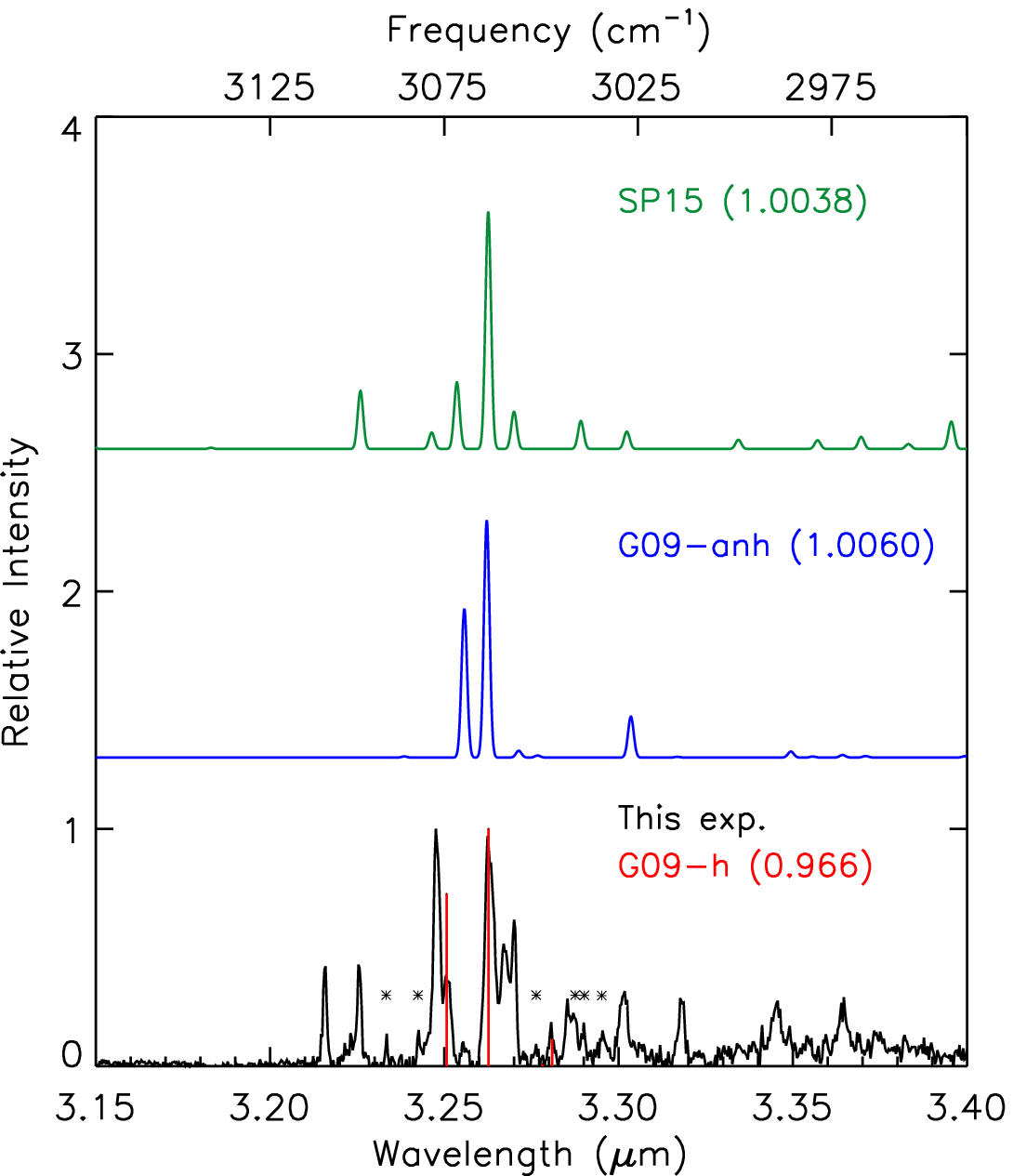}
\caption{The absorption spectra of naphthalene as predicted with (a) SP15, (b) G09-anh, and (c) G09-h together with the measured spectrum (This exp.). \label{fig1_na}}
\end{figure}

Figure \ref{fig1_na} displays the experimental and theoretical IR absorption spectra for napthalene. The spectrum shows narrow bands with widths varying between 1 and 3~cm$^{-1}$ and more than 16 well-separated bands of which a considerable fraction has intensities exceeding 20\% of the strongest band. In total, 23 experimental lines are identified and listed in Table \ref{table:lines}. Based on group theory, four IR-active transitions are present in the CH stretch region with either b$_{1u}$ or b$_{2u}$ symmetries\footnote{b$_{3u}$ transitions are IR-active, however none with this symmetry fall in the CH stretch region}. Indeed, G09-h predicts two intense bands that may be associated with the strongest and the two less intense bands that are of ambiguous assignment. A scaling factor of 0.966 was found to give the best agreement with experiment, in close agreement with \citet{Cane07}, considering that only the CH stretch region is considered here. 

Table \ref{table:lines} also shows the line positions (in cm$^{-1}$) reported in a previous study using cavity ring-down spectroscopy (CRDS) \citep{Hune04}. Most experimental bands are within 0.2~cm$^{-1}$. Some of the weaker bands deviate up to 1 ~cm$^{-1}$, which may be attributed to the inaccuracy with which the center can be determined. Our error in the positions is most likely smaller due to superior S/N ratio. We observe six additional bands not reported before (asterisks in Table \ref{table:lines} and Figure \ref{fig1_na}). Our spectra do not give evidence for the presence of the very weak bands reported before in \citet{Hune04} (3064.1, 3068.5, and 3098.9 ~cm$^{-1}$). The present study also shows narrower line widths, indicating lower internal temperatures. The improved resolution significantly changes the intensity distribution of the majority of the bands; e.g. the bands around 3100~cm$^{-1}$ show the smallest line widths (1.0 and 1.1~cm$^{-1}$, respectively) and have intensities of about 45\% of the intensity of the strongest band, whereas in the CRDS study the intensity of these bands was less than 20\% and displayed a broader structure. 

Figure  \ref{fig1_na} also shows the spectrum of naphthalene as predicted with G09-anh, convolved with 1 cm$^{-1}$ to resemble experimental resolution. A total of 11 double excitation bands are predicted, i.e., bands arising from $\Delta \nu_{i} = 2$ or $\Delta \nu{_i} = 1; \Delta \nu_{j} = 1$ transitions. The G09-anh frequencies are in very good agreement with experiment, the typical deviation of the strongest bands being only 0.6\%. This remaining deviation has been corrected for in the figure by a small scaling factor which also serves to facilitate easy comparison between observed and predicted spectra. Interestingly, none of the anharmonic bands have intensities comparable to those observed in experiment. This lack of intensity can arise from two possible scenarios. First, intensity is obtained from the anharmonicity of the potential energy and dipole moment surfaces along the relevant coordinates, making the $\Delta v$=1 selection rule no longer strictly valid. Second, intensity is acquired through Fermi coupling to the CH-stretch fundamentals, leading to intensity borrowing. 

To distinguish between these two, experiments on deuterated naphthalene were performed as the fundamental CD-stretch modes of naphthalene-d$_{8}$ are displaced to $\sim$4~$\mu$m while the overtones and combination bands originate from normal modes that are much less affected by deuteration. In the first scenario only slight shifts are expected, in the second case deuteration will lead to a dramatic reduction of intensity. We scanned a range of $3.12-3.85$ $\mu$m ($2600-3200$~cm$^{-1}$) taking into account the possible shifts of the combination bands due to deuteration, and found no signal. We thus conclude that the additional bands in naphthalene-h$_{8}$ derive their intensity from Fermi coupling to fundamental CH-stretch transitions. This means all bands are either of b$_{1u}$ or b$_{2u}$ symmetry with corresponding rotational contours; the envelop of b$_{2u}$ bands being narrower than for b$_{1u}$ bands. A rotational-contour analysis of the two strongest bands gives best agreement using a rotational temperature of 4~K and a homogeneous line width of 0.5~cm$^{-1}$. Indeed, two sets of band widths were measured; relatively large widths between 2.3-2.8~cm$^{-1}$ attributed to b$_{1u}$ transitions and narrower widths between 1.1-1.9~cm$^{-1}$ attributed to b$_{2u}$ transitions.

To properly include Fermi resonances, we calculated the spectrum of naphthalene using SP15. Like G09-anh, in total 11 IR-active combination bands are within the experimental range. The slight overestimation of the anharmonicity is even smaller, within 0.4\%. Unlike G09-anh, SP15 does give appropriate intensities and the agreement with experiment is improved although discrepancies remain. The major discrepancy concerns the number of observed bands. The calculations predict 15 bands while 23 bands are observed.  We therefore performed a combinatorial analysis based on experimental data. We determined the frequencies of all possible double combination bands using the sums of measured fundamental frequencies of naphthalene in the 6 $\mu$m region (see \citet{Hewe94} and references therein), including both IR and Raman active modes. On the basis of these modes (all CC stretches except for one CH bend) and the symmetry restrictions, 11 IR-active combination bands fall within a broad range of $3.15-3.5$ $\mu$m ($2900-3150$ cm$^{-1}$). Both calculation and combinatorial analysis are thus not able to account for all the observed bands. A similar exercise can be performed for triple combination bands, which provides a multitude of additional candidates with the correct symmetry that fall within the experimental range. Interestingly, we find that these dominantly involve CH bending modes, which are prone to coupling with CH stretching modes. We conclude that the 3~$\mu$m region of the absorption spectrum of naphthalene is dominated by Fermi resonances and that triple combination bands need to be taken into account as well. The latter would require incorporation of even higher-order couplings than currently included in the anharmonic analyses. A detailed assignment of the double combination bands in naphthalene as well as anthracene and tetracene ({\it vide infra}) will be presented in a separate study \citet{Mack15b}. Also, methods that incorporate higher-order terms are presently under investigation and will be presented in future work.

\subsection{Anthracene (C$_{14}$H$_{10}$)}

\begin{figure}
\plotone{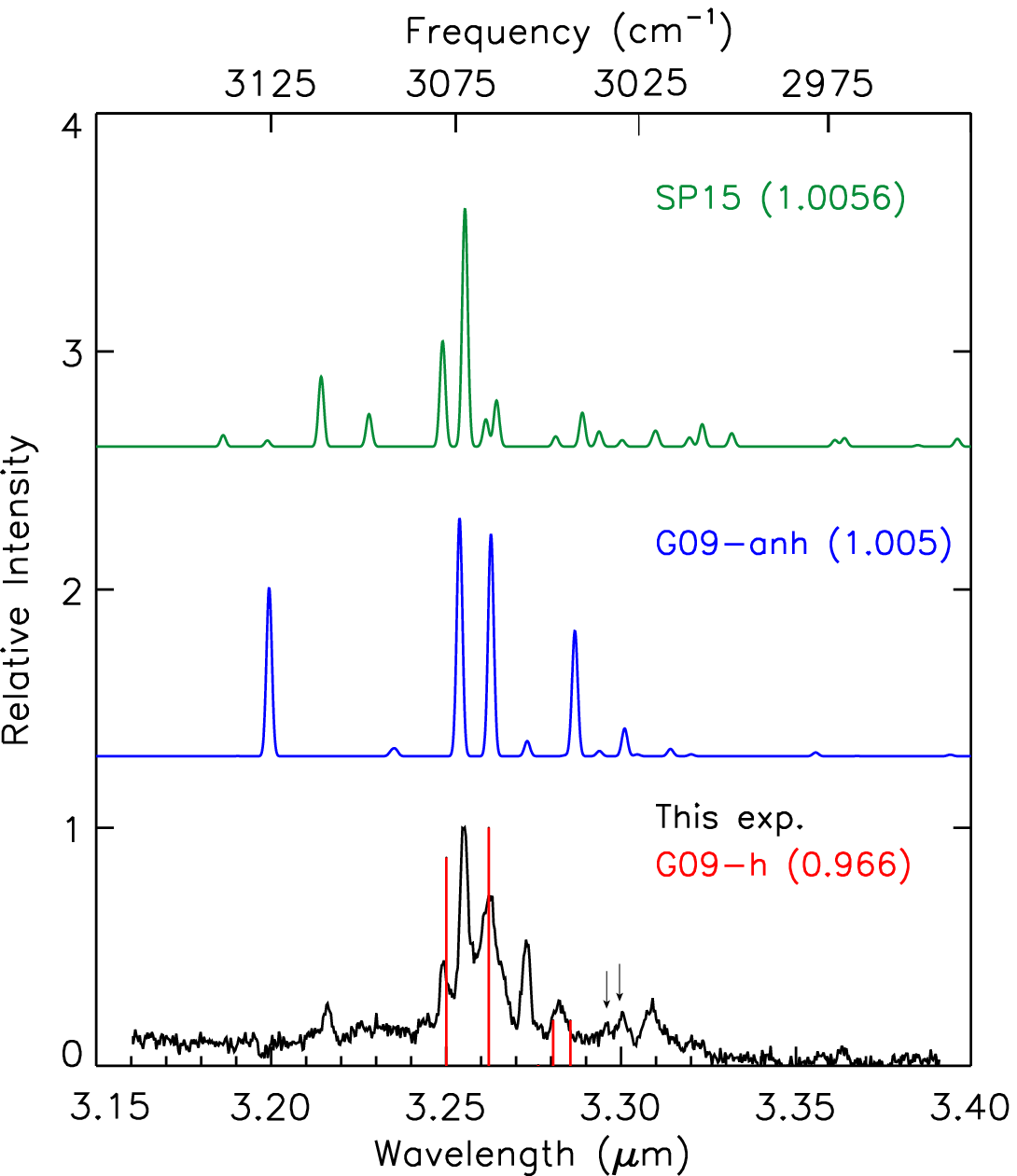}
\caption{The absorption spectra of anthracene as predicted with (a) SP15, (b) G09-anh, and (c) G09-h together with the measured spectrum (This exp.). \label{fig2_an}}
\end{figure}

The experimental and theoretical absorption spectra of anthracene are shown in Figure \ref{fig2_an}. The positions of the observed bands are listed in table \ref{table:lines}. Again, the experimental spectrum reveals more bands with appreciable intensities than predicted by harmonic calculations, which gives best agreement with a scaling factor of 0.966. The G09-anh spectrum again slightly overestimates the anharmonicity requiring a scaling factor less than 0.5\% to overlap the strongest bands. As before, G09-anh predicts many additional bands with low to zero intensity. Interestingly, the band at 3.2 $\mu$m is much higher than observed in experiment and might be caused by a missing resonance. Indeed, incorporating Fermi coupling using SP15 permits to redistribute intensity over multiple transitions, which not only leads to a lower intensity band at 3.22 $\mu$m but also to better overall agreement with experiment.

Table \ref{table:lines} shows the comparison to the previous CRDS study \citep[]{Hune04}. We measure seven more bands, which can be attributed to improved cooling conditions. All other band positions are within experimental error with the exception of the previously reported band at 3032.1 ~cm$^{-1}$. This band consists of two close-lying bands at 3030.0 and 3033.7 ~cm$^{-1}$ in our experiment (arrows in figure \ref{fig2_an}). Bands with similar widths as in naphthalene are observed indicating similar cooling conditions ($\sim$2.5~cm$^{-1}$). Simulation of the rotational contours that can be expected for sensible rotational temperatures under the present experimental conditions show $-$ in contrast to what was observed for naphthalene $-$ that the widths of the observed bands are not determined by their rotational contours. Instead, one has to conclude that these bands consist of several overlapping bands, and that a combinatorial analysis of the entire spectrum as was done for naphthalene is not possible. Moreover, the state density in anthracene at these energies is already so large that this approach would likely not lead to a conclusive assignment.

\subsection{Tetracene (C$_{18}$H$_{12}$)}

\begin{figure}
\plotone{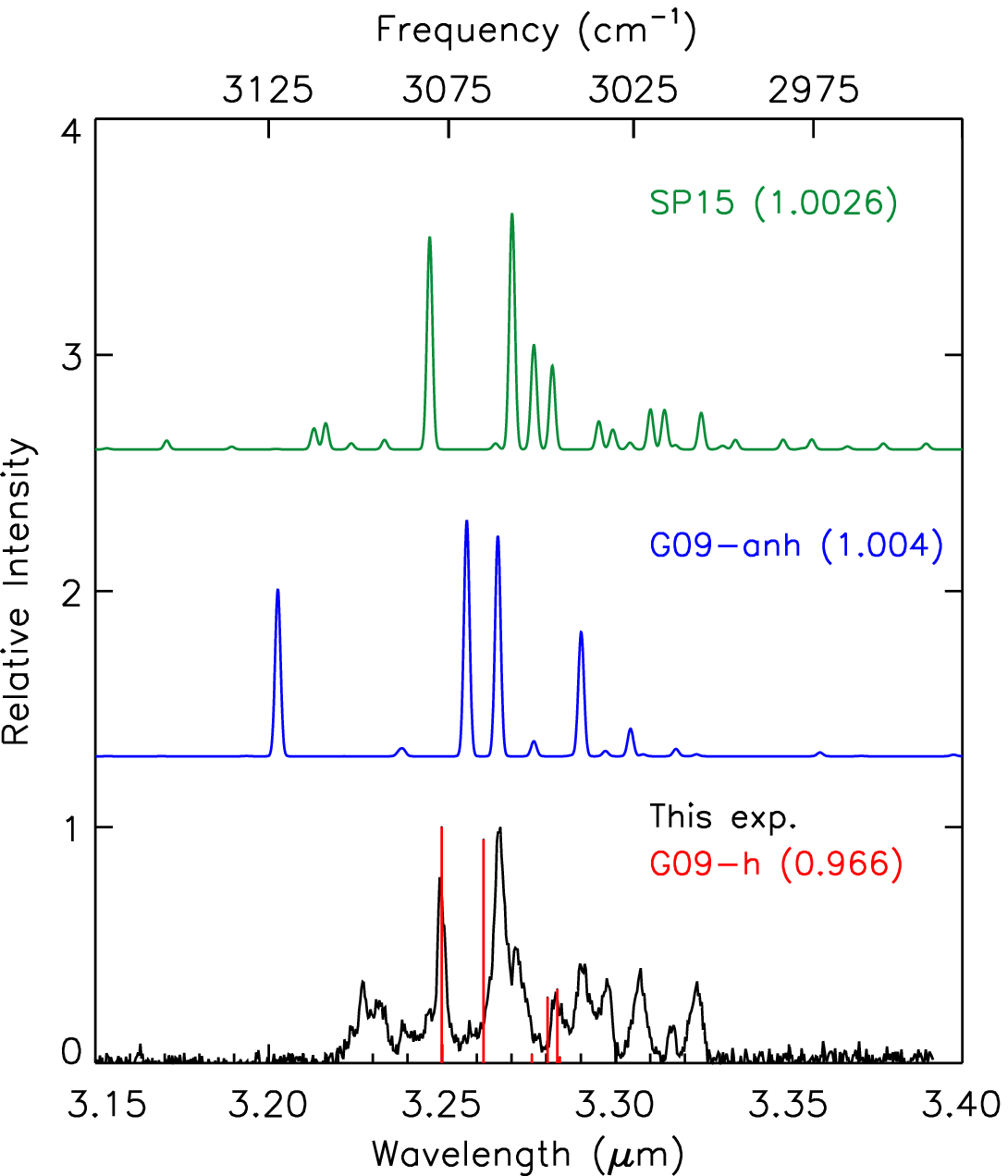}
\caption{The absorption spectra of tetracene as predicted with (a) SP15, (b) G09-anh, and (c) G09-h together with the measured spectrum (This exp.). \label{fig3_te}}
\end{figure}

Figure \ref{fig3_te} shows the experimental and theoretical absorption spectra of tetracene. We observe more than 19 well-resolved bands with line widths between $2.3-4.4$~cm$^{-1}$ (table \ref{table:lines}). As for anthracene, it is most likely that the width of these bands is not determined by the rotational contour of one specific transition, but results from the overlap of several unresolved transitions. The G09-h calculation predicts that only four of the twelve symmetry-allowed CH stretch bands have an appreciable intensity. For these four bands, a rather poor agreement with experiment is observed, in contrast to naphthalene and anthracene where a reasonably good agreement could be obtained for the fundamental bands after scaling. The G09-anh calculation, on the other hand, performs much better in this case; its frequencies are in close agreement with the experimentally observed strong bands. When including the Fermi coupling with SP15, even better agreement can be achieved.

\begin{figure}
\includegraphics[scale=0.80]{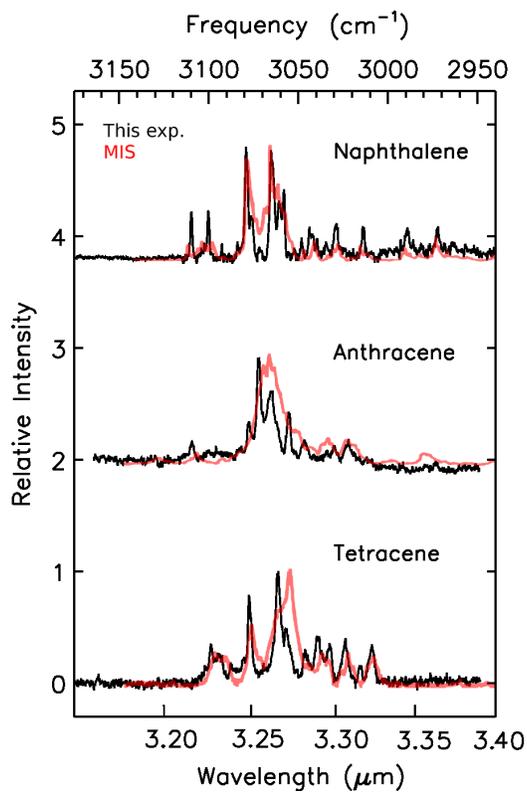}
\caption{The absorption spectra of (a) naphthalene, (b) anthracene, and (c) tetracene (This exp.) compared to matrix isolation spectroscopy measurements (MIS) \citep{Hudg98a}. \label{fig4_mis}}
\end{figure}

Since for tetracene no gas-phase IR absorption spectra have been reported before, we compare the present data with data obtained in Ar matrix isolation studies (MIS) \citep[]{Hudg98a}. Figure \ref{fig4_mis} shows all measured spectra compared to their MIS counterparts. The agreement in all three cases is very high with superior S/N ratio and cooling in the present study, leading to narrower and more resolved lines. As expected from matrix-induced effects \citep{Jobl94}, relatively small spectral shifts are observed for most bands. The larger shifts are most probably due to a redistribution of relative intensities; e.g. for tetracene, an apparent large shift (of 6.1~cm$^{-1}$) seems to occur for the most intense band, however, it is more likely that the relative intensities of the double structure around 3.27 $\mu$m are different for the MIS spectrum, leading to another most intense band.

\section{Astrophysical implications}

The present results show that harmonic DFT calculations fail dramatically in predicting high-resolution experimental absorption spectra of PAHs in the 3-$\mu$m region. This is important since such calculations are routinely used to help in the interpretation of astronomical observations, and in particular, in efforts to shed light on the evolution of the interstellar carbon inventory. In the past, the discrepancy between frequencies calculated in the harmonic approximation and MIS spectra have commonly been overcome by introducing a scaling factor of 0.966, corresponding to a shift in the order of 100 cm$^{-1}$. In this context, five key observations are made in the present study.

Firstly, MIS spectra agree well with the gas phase spectra measured here in terms of the position of the bands. However, there are differences in relative intensities, reflecting the importance of Fermi resonances (which act differently in a matrix then in the gas phase), and this may lead to subtle shifts in the intensity averaged peak position \citep{Hune04}.

Secondly, for the strongest bands in the spectrum, the anharmonic predictions obtained using SP15 all fall within 0.5\% from the experimental value, significantly reducing the error introduced by the use of empirical scaling factors. 

Thirdly, in order to ``translate" spectroscopic data obtained from calculations or low-temperature experiments into emission spectra useful for comparison with observational data, a typical empirical redshift of 15 cm$^{-1}$ is commonly applied \citep{Baus10}. As demonstrated by the pioneering study on the low-frequency modes of naphthalene \citep{Pira09}, mode-specific shifts in emission spectra are more accurately predictable as our knowledge of the anharmonic potential expands. We are presently working on such modelling efforts.

Fourthly, Fermi resonances may contribute significantly to the profile and structure of the 3-$\mu$m band. The frequencies of the major CH-stretch fundamental bands are within a spread of around 50 cm$^{-1}$ about the same for all types of PAHs \citep{Hudg98a,Hudg98b}. The frequencies of the modes that are potentially involved in combination bands, on the other hand, are much more sensitive to the finer details of the structure. These observations match astronomical studies of IR emission that show a very prominent emission band at 3.29 $\mu$m corresponding to the 1-0 transition of the fundamental CH-stretch bands, and a broad plateau in the $3.1-3.7$ $\mu$m region indicated as the 3-$\mu$m plateau or vibrational quasi-continuum \citep{ATB89,Geba89}. DFT calculations so far have not been able to explain and characterize this plateau. Our studies demonstrate that anharmonic couplings can be responsible for some of the structure observed on the main band as well as the wings on both the blue and the red side of the main astronomical band.

Fifthly, as our study demonstrates anharmonicity and Fermi resonances are particularly important in the 3 um region of the spectrum. An assessment of their influence for the other bands remains to be determined.

It may be noticed that the spectral range over which features are observed when comparing the three experimental spectra seems to become narrower with increasing PAH size. This aspect is further discussed in separate studies on the size and structure dependence of the 3-$\mu$m region \citep{Mack15b,Mack15c,Malt15}.

\section{Conclusions}
We report on the absorption spectra of linear PAHs in the 3-$\mu$m region using UV-IR double resonance laser spectroscopy under cold and isolated conditions. Efficient cooling and excellent S/N ratios lead to well-resolved spectra that show a plethora of vibrational transitions. Comparison with harmonic predictions show that the fraction of intensity that is not associated with fundamental transitions may easily exceed 50\%, and therefore cannot be neglected. A detailed analysis of the absorption spectrum of naphthalene has demonstrated that the additional activity originates from vibrational coupling of the bath of `dark' states with the intensity-carrying `bright' states, and not purely from anharmonic effects. The present studies emphasize the necessity of properly incorporating Fermi resonances and higher-order vibrational couplings. First results incorporating Fermi resonances with SPECTRO are rather promising, as they indeed seem to lead to a qualitatively correct description of the intensity distribution. However, the measured spectra show more bands than the number of possible double combination bands. This indicates that incorporation of higher-order couplings, i.e., triple combination bands involving CH bending modes, may be necessary to obtain a proper quantitative description.

Detailed studies of the spectra of PAHs are very timely. The launch of the James Webb Space Telescope will open up a revolutionary window on the PAH spectrum with unprecedented spatial and spectral resolution. Studies on larger PAHs are needed to determine in more detail how the size of a PAH affects the intensity distribution over the 3-$\mu$m band. Analogous studies on molecules with the same number of rings but with a different structure -leading among others to the presence of bay and non-bay hydrogen atoms - is also of considerable interest: if a relation can be derived between the structure of the molecule and the shape and position of the 3-$\mu$m band, progress could be made towards determining the chemical composition of interstellar objects \citep{Cand12}. Such IR absorption studies on larger PAHs and on PAHs with different structures are presently being performed in our lab \citep{Malt15}.

\acknowledgments
The experimental work was supported by The Netherlands Organization for Scientific Research (NWO). Studies of interstellar PAHs at Leiden Observatory have been supported through the advanced European Research Council Grant 246976 and a Spinoza award. Computing time has been made available by NWO Exacte Wetenschappen (project MP-270-13 and MP-264-14) and calculations were performed at the LISA Linux cluster of the SurfSARA supercomputer center in Almere, The Netherlands. XH and TJL gratefully acknowledge support from the NASA 12-APRA12-0107 grant. XH acknowledges the support from NASA/SETI Co-op Agreement NNX15AF45A.

\clearpage

\begin{table}
\begin{center}
\caption{The measured bands of naphthalene, anthracene, and tertracene with frequencies in cm$^{-1}$ and intensities normalised to the respective strongest transitions. Comparison is made to previous gas-phase frequencies (cm$^{-1}$) recorded with cavity ring down spectroscopy (CRDS) \citep{Hune04}. \label{table:lines}}
\begin{tabular}{lccrccccc}
\tableline\tableline
	\multicolumn{4}{c}{naphthalene (C$_{10}$H$_8$)}				
	& \multicolumn{3}{c}{anthracene (C$_{14}$H$_{10}$)}		
	& \multicolumn{2}{c}{tetracene (C$_{18}$H$_{12}$)} 				\\
\cline{1-4} 	\cline{8-9}
	\multicolumn{3}{c}{this work}	& CRDS				
	& \multicolumn{2}{c}{this work}	& CRDS		
	& \multicolumn{2}{c}{this work} 				\\
	freq. 		& rel. int.	& symm.	& freq.	
	& freq. 	& rel. int.	& freq.
	& freq. 	& rel. int.					\\
\tableline
\tableline
	2963.8	& 0.2			& 		& 2965.3	& 2973.2	& 0.10	& 		& 3008.9	& 0.39	\\
	2972.4	& 0.3			& b1u	& 2973.1	& 2979.6	& 0.08	& 		& 3015.3	& 0.2		\\
	2981.3	& 0.2			& 		& 2980.9	& 2992.5	& 0.07	& 		& 3023.7	& 0.44	\\
	2989.0	& 0.28		& b1u	& 2989.1	& 3011.8	& 0.15	& 		& 3032.6	& 0.4		\\
	3014.0	& 0.3			& b2u	& 3013.7	& 3022.0	& 0.30	& 3021.7	& 3038.5	& 0.45	\\
	3029.0	& 0.31		& b1u	& 3029.1	& 3030.0	& 0.25	& 		& 3039.7	& 0.46	\\
	3034.5\tablenotemark{*}	& 0.15	& b1u	& -		& 3033.7	& 0.20	& 3032.1	& 3046.2	& 0.34	\\
	3039.5\tablenotemark{*}	& 0.19	& b2u	& -		& 3046.7	& 0.30	& 3047.5	& 3050.6	& 0.18	\\
	3042.3$^{(*)}$	& 0.29	& b1u	& \multirow{2}{*}{\big \rangle 3043.7} 
							    			& 3055.4	& 0.54	& 		& 3054.8	& 0.38	\\
	3043.8$^{(*)}$	& 0.35	& b2u	& 		& 3062.3	& 0.45	& 		& 3056.8	& 0.51	\\
	3048.2	& 0.19		& b2u	& 3048.9	& 3065.3	& 0.72	& 3064.3	& 3061.1	& 1		\\
	3052.2\tablenotemark{*}	& 0.17	& 		& -		& 3066.9	& 0.64	& 3067.5	& 3066.6	& 0.2		\\
	3058.1	& 0.65		& b2u	& 3057.9	& 3071.9	& 1.00	& 3072.5	& 3069.5	& 0.23	\\
	3060.5	& 0.54		& b1u	& 3061.1	& 3077.8	& 0.40	& 3077.9	& 3077.6	& 0.8		\\
	3065.2	& 0.99		& b1u	& 3065.1	& 3081.8	& 0.24	& 		& 3080.4	& 0.27	\\
	3071.4	& 0.1			& b2u	& 3069.9	& 3095.9	& 0.24	& 		& 3087.9	& 0.23	\\
	3076.2	& 0.37		& b2u	& 3074.1	& 3109.6	& 0.32	& 3109.9	& 3094.1	& 0.31	\\
	3079.2	& 1			& b2u	& 3079.1	& 		& 		& 		& 3098.8	& 0.38	\\
	3083.9\tablenotemark{*}	& 0.16	& b2u	& -		& 		& 		& 		& 3101.5	& 0.19	\\
	3092.6\tablenotemark{*}	& 0.21	& 		& -		& 		& 		& 		& 		& 		\\
	3100.2	& 0.45		& b2u	& 3100.1	& 		& 		& 		& 		& 		\\
	3102.6	& 0.21		& 		& 3102.2	& 		& 		& 		& 		& 		\\
	3109.4	& 0.45		& b2u	& 3109.3	& 		& 		& 		& 		& 		\\
\tableline
\tablenotetext{*}{ These are the newly measured lines. The brackets denote that these two lines were previously measured as one.}
\end{tabular}
\end{center}
\end{table}

\clearpage

\clearpage

\clearpage

\clearpage

\end{document}